\documentclass[aps,prc,twocolumn,showpacs,groupedaddress,floatfix]{revtex4}

\usepackage{graphicx}
\usepackage{dcolumn}
\usepackage{amsfonts}

\begin{document}

\title{Heavy-ion interaction potential deduced from density-constrained\\
        time-dependent Hartree-Fock calculation}

\author{A.S. Umar and V.E. Oberacker}
\affiliation{Department of Physics and Astronomy, Vanderbilt University, Nashville, Tennessee 37235, USA}

\date{\today}


\begin{abstract}
We present a new method for calculating the heavy-ion interaction potential
from a density-constrained time-dependent Hartree-Fock calculation.
\end{abstract}
\maketitle


The study of internuclear potentials for heavy-ion collisions is
of fundamental importance for the formation of
superheavy elements and nuclei far from stability.
While asymptotically such potentials are determined from
Coulomb and centrifugal interactions, the short distance
behavior strongly depends on the nuclear surface properties
and the readjustments of the combined nuclear system,
resulting in potential pockets, which determine the
characteristics of the compound nuclear system.

Among the various approaches for calculating ion-ion
potentials are:
1) Phenomenological models such as the Bass model~\cite{Ba74,Ba80},
the proximity potential~\cite{BR77,RV78,SG84,BH78}, and potentials obtained via
the double-folding method~\cite{SL79,BB77,RO83a,RO83b}. Some of these potentials
have been fitted to experimental fusion barrier heights
and have been remarkably successful in describing scattering data.
2) Semi-microscopic and full microscopic calculations such as the
macroscopic-microscopic method~\cite{KN79,MS04,II05}, the 
asymmetric two-center shell-model~\cite{MG72}, 
constrained Hartree-Fock (CHF) with a constraint on the quadrupole moment
or some other definition of the internuclear distance~\cite{ZM76,BB04},
and other mean-field based calculations~\cite{DN02,DP03,GB05}.

One common physical assumption used in many of the semi-microscopic
calculations is the use of the {\it frozen density} or the {\it sudden}
approximation. As the name suggests, in this approximation the nuclear densities
are unchanged during the computation of the ion-ion potential as a function
of the internuclear distance. On the other hand, the microscopic calculations
follow a minimum energy path and allow for the rearrangement of the nuclear
densities as the relevant collective parameter changes.
As it was pointed out in Ref.~\cite{MS04}, CHF calculations seldom
produce the correct saddle-point since the system can follow any
one of the minimum potential valleys in the multi-dimensional potential
energy surface.
In this paper, we
shall call this the {\it static adiabatic approximation} since a real
adiabatic calculation would involve a fully dynamical calculation, thus
also including the effects of dynamical rearrangements.

One conclusion that may be reached from the discussion above is that ultimately
we would like to have an approach for calculating internuclear potentials
which is time-dependent and is unrestricted in the choice of collective
variables. In this paper we provide such an approach in which time-dependent
Hartree-Fock (TDHF) is used for the nuclear dynamics and the potential
energy is calculated by constraining the time-dependent density.

The {\it density constraint} is a novel numerical method that was developed in
the mid 1980's~\cite{CR85,US85} and was used to provide a microscopic
description of the formation of shape resonances in light systems~\cite{US85}.
In this approach the TDHF time-evolution takes place with no restrictions.
At certain times during the evolution the instantaneous density is used to
perform a static Hartree-Fock minimization while holding the total density constrained
to be the instantaneous TDHF density. In essence, this provides us with the
TDHF dynamical path in relation to the multi-dimensional static energy surface
of the combined nuclear system. Since we are constraining the total density
all moments are simultaneously constrained. In the traditional CHF notation
this corresponds to the replacement
\begin{equation}
\lambda\hat Q \longrightarrow \lambda\hat \rho\;.
\end{equation}
The numerical procedure for implementing this constraint and the method for
steering the solution to $\rho_{\mathrm{TDHF}}(\mathbf{r},t)$ is discussed in Refs.~\cite{CR85,US85}.
The convergence property is as good if not better than in the traditional CHF
calculations with a constraint on a single collective degree of freedom.

In this paper, we shall call the energy of the system obtained by the density constraint
method, $E_{DC}(R)$, where the dependence is on the instantaneous internuclear separation, $R(t)$.
Since this quantity contains no translational kinetic
energy (taken out by the static minimization) it is actually a potential as we show
below. We define the excitation energy of the system as
\begin{equation}
E^{\text{*}}(R)=E_{\mathrm{TDHF}}-T_{R}-E_{\mathrm{DC}}(R)\;,
\label{eq:ex}
\end{equation}
where $E_{\mathrm{TDHF}}$ is the total TDHF energy
\begin{equation}
E_{\mathrm{TDHF}}=\int d^{3}r\text{ }{\cal H}(\mathbf{r},t)\;,
\end{equation}
which is conserved throughout the calculation, and $T_{R}$ is the
instantaneous translational energy between the two nuclei
\begin{equation}
T_{R}=\frac{1}{2}\mu {\dot{R}}^{2}\;,
\end{equation}
with $\mu$ being the reduced mass of the system, and $\dot{R}$ is the
velocity associated with the internuclear separation coordinate $R(t)$.
At the same time the total TDHF energy can be written in terms of the
excitation energy as
\begin{equation}
E_{\mathrm{TDHF}}=T_{R}+V(R)+E^{\text{*}}(R)\;,
\end{equation}
such that when combined with Eq.~(\ref{eq:ex}) shows that $V(R)=E_{\mathrm{DC}}(R)$.
However, the density constrained potential still contains the binding energies of
individual nuclei, which should be subtracted out;
\begin{equation}
V(R)\rightarrow V(R)=E_{\mathrm{DC}}(R)-E_{A_{1}}-E_{A_{2}}\;.
\label{eq:vr}
\end{equation}
Eq.~(\ref{eq:vr}) is the internuclear potential and contains
{\it no free parameters}. Given an effective nuclear interaction, such
as the Skyrme force,  $V(R)$
can be constructed by performing a TDHF evolution and minimizing
the energy at certain times to obtain
$E_{\mathrm{DC}}(R)$, while $E_{A_{1}}$ and $E_{A_{2}}$ are the
results of a static Hartree-Fock calculation with the same effective
interaction. One can see that the expression also has the correct
asymptotic behavior since numerically for large $R$ we exactly get
\begin{equation}
E_{\mathrm{DC}}(R_{\mathrm{max}})=E_{A_{1}}+E_{A_{2}}+\frac{Z_{1}Z_{2}e^{2}}{R_{\mathrm{max}}}\;,
\end{equation}
such that
\begin{equation}
V(R_{\mathrm{max}})=\frac{Z_{1}Z_{2}e^{2}}{R_{\mathrm{max}}}\;,
\end{equation}
so that normalization of $V(R)$ is not necessary.
\begin{figure}[!htb]
\begin{center}
\includegraphics*[scale=0.30]{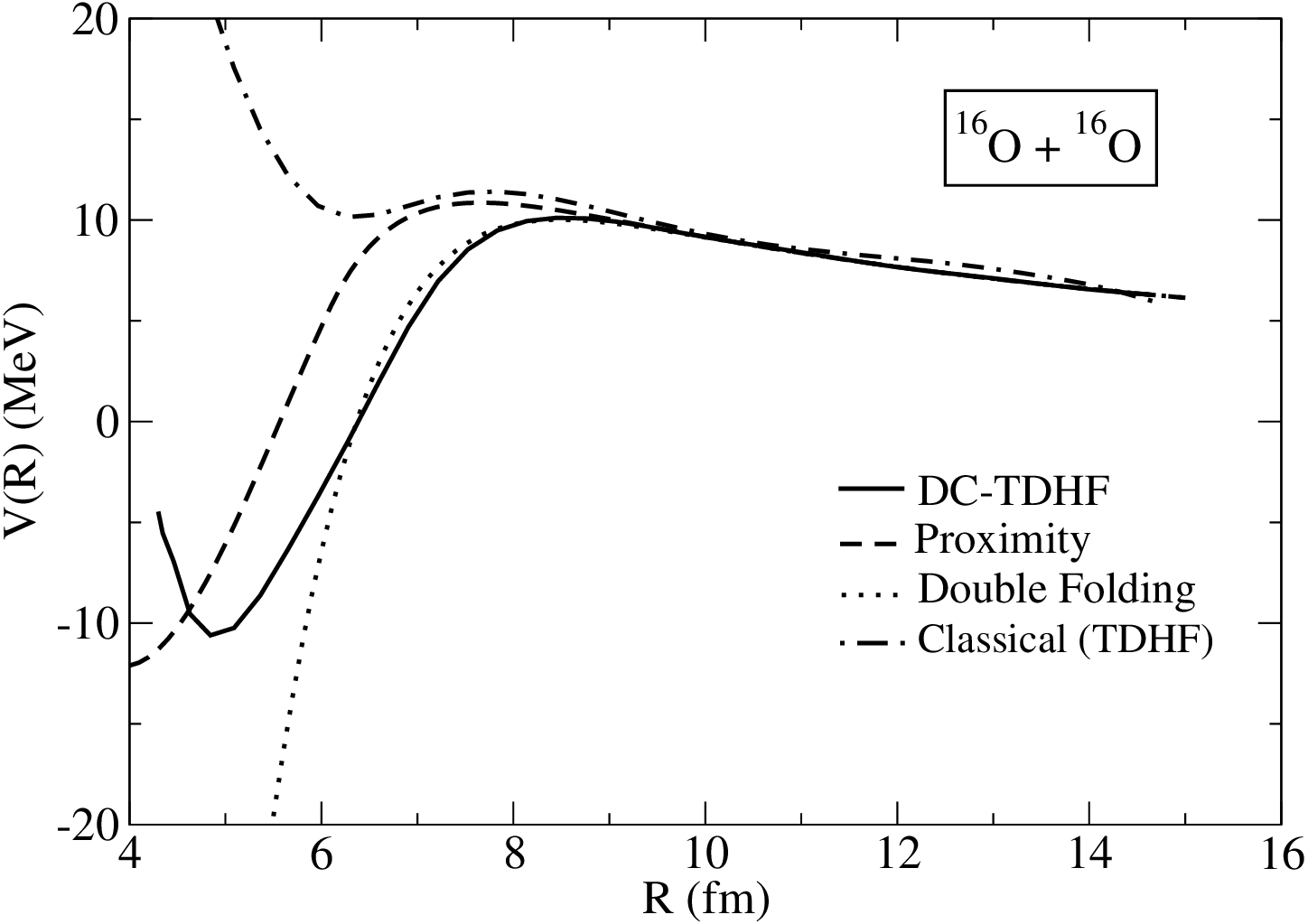}
\caption{\label{fig:OO} Internuclear potential obtained from Eq.~\protect\ref{eq:vr}
and various model calculations for the head-on collision of the $^{16}$O+$^{16}$O system
at $E_{\mathrm{c.m.}}=34$~MeV.}
\end{center}
\end{figure}

We have carried out a number of TDHF calculations with accompanying
density constraint calculations to compute $V(R)$ given by Eq.~(\ref{eq:vr}).
A detailed description of our new three-dimensional unrestricted TDHF code
has recently been published in Ref~\cite{UO06}.
For the effective interaction we have used the Skyrme SLy5 force~\cite{CB98}
including all of the time-odd terms. In Fig.~\ref{fig:OO} we show the result of
our calculation for the head-on (zero impact parameter) collision of $^{16}$O+$^{16}$O
at $E_{\mathrm{c.m.}}=34$~MeV.
Also shown in Fig.~\ref{fig:OO} are two widely used phenomenological potentials, the
standard  proximity potential for two spherical nuclei~\cite{BR77,RV78,SG84} 
and the double-folding potential with M3Y effective NN interaction~\cite{SL79,BB77,RO83a,RO83b}.
We evaluate the double-folding integral for the strong nuclear and
Coulomb interaction in momentum space~\cite{RO83b}. For the charge and
matter densities  we utilized generalized Fermi distributions
whose parameters were determined from electron scattering experiments~\cite{SM70}.
The double-folding potential agrees almost perfectly with the DC-TDHF approach
for distances $R \geq 6$ fm. At smaller distances, the double-folding potential tends
to overestimate the nuclear interaction as a result of the (unphysical) frozen density approximation.
The classical TDHF curve corresponds to the older definition
\begin{equation}
V(R)=E_{\mathrm{c.m.}}-T_{R}\;,
\end{equation}
used in Ref.~\cite{KD77}.

We have also repeated the above calculations for different center-of-mass energies.
In this case, we find that the results at the barrier do not appreciably change
but the depth of the potential increases for lower energies. For $E_{\mathrm{c.m.}}=12$~MeV
the potential is about 1.5~MeV deeper than the one at 34~MeV. One comment is
required regarding the calculation of the internuclear separation $R$. As usual,
this quantity becomes somewhat unclear for a strongly overlapping system. In
our case we use the standard TDHF approach of finding left and right dividing
planes and computing the centers of the density in these two halves and thus
the separation.
\begin{figure}[!htb]
\begin{center}
\includegraphics*[scale=0.30]{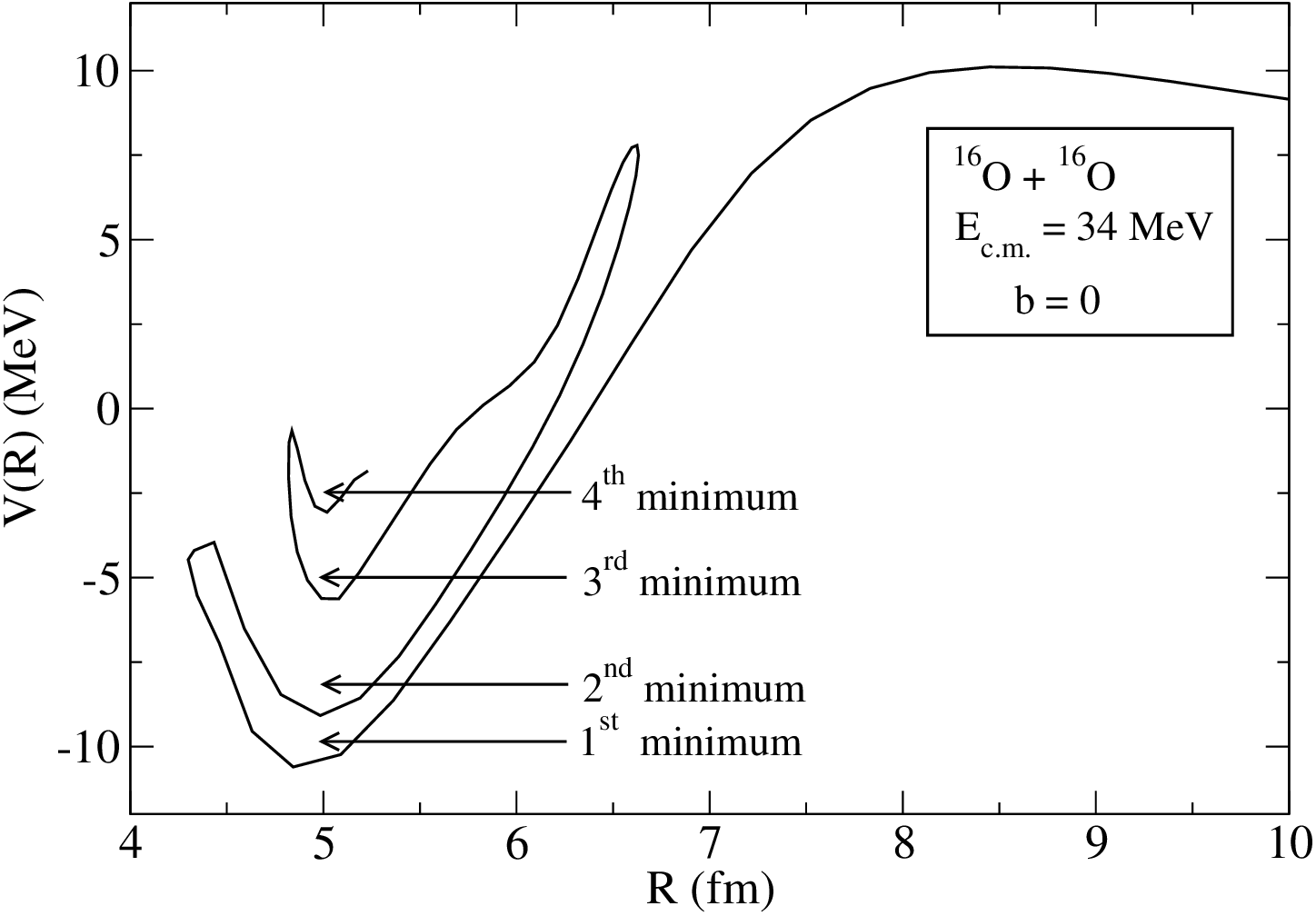}
\caption{\label{fig:ladder} Internuclear potential obtained from Eq.~\protect\ref{eq:vr}
shown for the entire time evolution of the $^{16}$O+$^{16}$O system at $E_{\mathrm{c.m.}}=34$~MeV.}
\end{center}
\end{figure}

In TDHF, fusion occurs when the relative kinetic
energy in the entrance channel is entirely converted into internal
excitations of a single well defined compound nucleus.
The dissipation of the relative kinetic energy into internal excitations is
due to the collisions of the nucleons with the ``walls'' of the
self-consistent mean-field potential. TDHF studies demonstrate that the
randomization of the single-particle motion occurs through repeated
exchange of
nucleons from one nucleus into the other. Consequently, the equilibration of
excitations is very slow and it is sensitive to the details of the
evolution of the shape of the composite system.
This is in contrast to
most classical pictures of nuclear fusion, which generally assume near
instantaneous, isotropic equilibration. This equilibration can be observed
in the DC-TDHF approach by tracking the evolution of the excitation energy
in time for a system on the way to fusion, or alternately one can examine
the change in the internuclear potential for the compound system. In Fig.~\ref{fig:ladder}
we show this for the $^{16}$O+$^{16}$O system corresponding to the case shown
in Fig.~\ref{fig:OO}.
After passing the first minimum the system falls back to a second minimum and
climbs up the potential barrier, but it cannot overcome the barrier due to some
of the energy being converted into internal excitations, consequently it falls back
to a third minimum, fourth minimum, and so on, until complete equilibration.
This {\it potential ladder} effect is characteristic for all fusing systems.
\begin{figure}[!htb]
\begin{center}
\includegraphics*[scale=0.30]{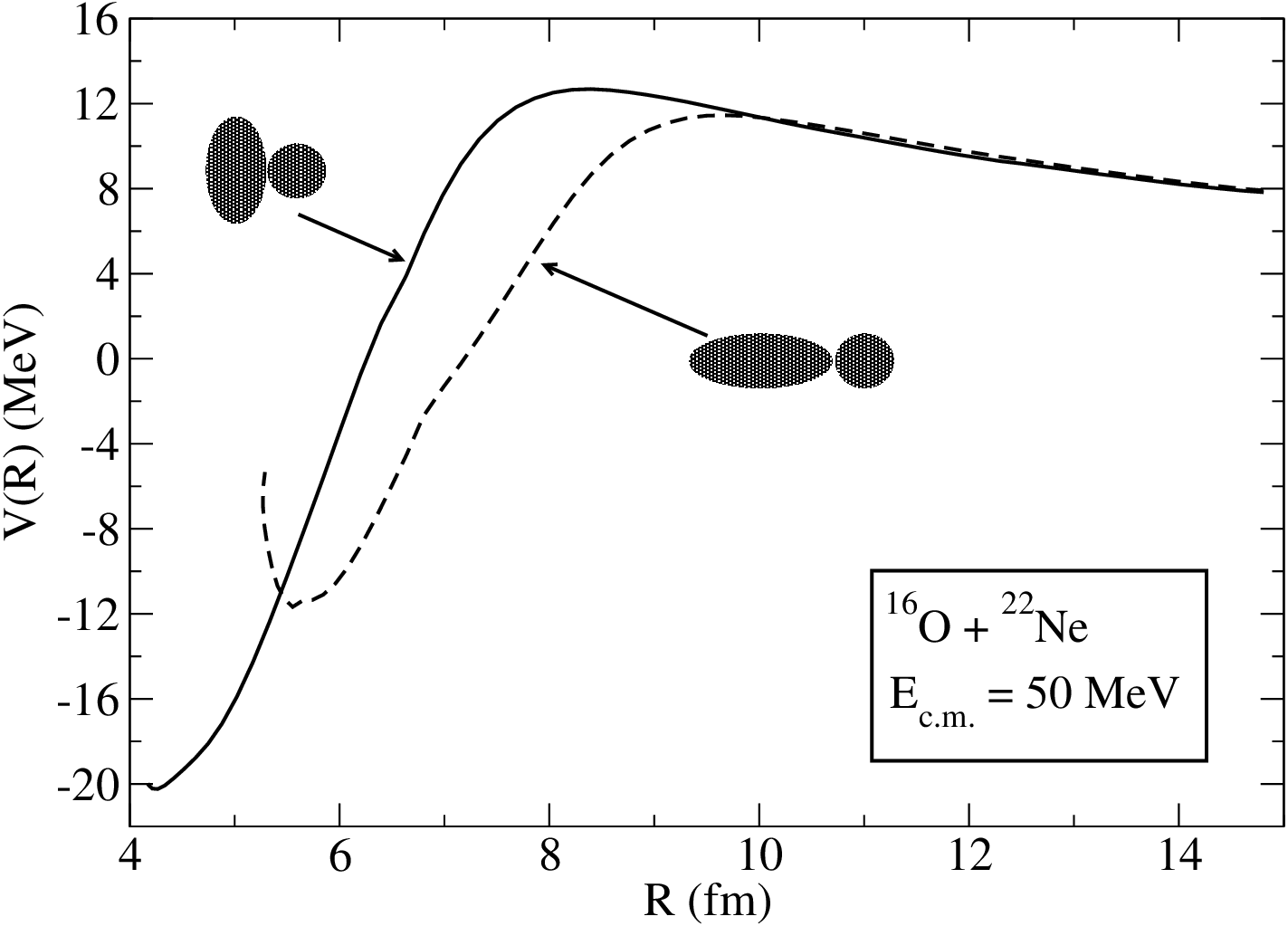}
\caption{\label{fig:align} Internuclear potential obtained from Eq.~\protect\ref{eq:vr}
shown for the evolution of the $^{16}$O+$^{22}$Ne system at $E_{\mathrm{c.m.}}=50$~MeV.
The two curves correspond to different orientations of the Ne nucleus.}
\end{center}
\end{figure}

Finally, we have performed calculations for the $^{16}$O+$^{22}$Ne system at
$E_{\mathrm{c.m.}}=50$~MeV. In this case the $^{22}$Ne nucleus shows a strong
axial deformation, which can have different orientations with respect to the collision axis.
We have recently reported a procedure for doing such calculations within the TDHF framework
in Ref.~\cite{UO06b}. Here, we show the result of our potential calculations for two orientations
of the $^{22}$Ne nucleus, one for which the symmetry axis of the Ne is aligned with the
collision axis and the other for which the symmetry axis is perpendicular to the collision
axis. In TDHF, for the first case (aligned with collision axis) we see no fusion at this
energy, whereas for the perpendicular alignment the system fuses. In Fig.~\ref{fig:align}
we show the results of our potential calculations. We observe that the barrier height
and its position in $R$ space, and the potential minimum and its position are considerably
different for the two orientations.

In summary, we have presented a method for calculating internuclear potentials directly from the
TDHF time-evolution of the colliding system. The method uses the {\it density
constraint} to trace the TDHF trajectory in relation to the static multi-dimensional
energy surface of the combined system. Since the TDHF evolution is unhindered all of
the collective dynamics associated with the evolution are included in the
calculations. We believe this provides a unique way to calculate ion-ion potentials
from mean-field calculations. Of course, we can only perform such calculations
for energies above the barrier since TDHF is a semi-classical theory in this regard.
Finally, the results are expected to be only as good as the TDHF description of the
particular system under study.

This work has been supported by the U.S. Department of Energy under grant No.
DE-FG02-96ER40963 with Vanderbilt University. Some of the numerical calculations
were carried out at the IBM-RS/6000 SP supercomputer of the National Energy Research
Scientific Computing Center which is supported by the Office of Science of the
U.S. Department of Energy.



\end{document}